\documentstyle[aps,preprint,epsfig]{revtex}

\begin{document}
\tighten
\hfuzz=4pt

\preprint{\font\fortssbx=cmssbx10 scaled \magstep2
\hbox to \hsize{
\hbox{\fortssbx University of Wisconsin - Madison}
\hfill$\vcenter{\tighten
                \hbox{\bf MADPH-99-1135}
                \hbox{\bf IFT-P.064/99}
                \hbox{August 1999}}$}}

\title{\vspace*{.35in}
Color Evaporation Induced Rapidity Gaps}

\author{O.\ J.\ P.\ \'Eboli, E.\ M.\ Gregores}
\address{Instituto de F\'{\i}sica Te\'orica,
Universidade Estadual Paulista \\
Rua Pamplona 145, 01405-900, S\~ao Paulo -- SP, Brazil}

\author{and\\ F.\ Halzen}
\address{Department of Physics, University of Wisconsin \\
Madison, WI 53706, USA}

\maketitle

\thispagestyle{empty}

\begin{abstract}
We show that soft color rearrangement of final states can account for the
appearance of rapidity gaps between jets.  In the color evaporation model the
probability to form a gap is simply determined by the color multiplicity of
the final state. This model has no free parameters and reproduces all data
obtained by the ZEUS, H1, D\O, and CDF collaborations.

\end{abstract}

\draft

\newpage

\section{Introduction}

We show that the appearance of rapidity gaps between jets, observed at the
HERA and Tevatron colliders, can be explained by supplementing the string
model with the idea of color evaporation, or soft color. The inclusion of soft
color interactions between the dynamical partons, which rearranges the string
structure of the interaction, leads to a parameter-free calculation of the
formation rate of rapidity gaps.  The idea is extremely simple. Like in the
string model, the dynamical partons are those producing the hard interactions,
and the left-over spectators.  A rapidity gap occurs whenever final state
partons form color singlet clusters separated in rapidity.  As the partons
propagate within the hadronic medium, they exchange soft gluons which modify
the string configuration.  These large-distance fluctuations are probably
complex enough for the occupation of different color states to approximately
respect statistical counting.  The probability to form a rapidity gap is then
determined by the color multiplicity of the final states formed by the
dynamical partons, and nothing else. All data obtained by ZEUS \cite{zeus}, H1
\cite{h1}, D\O\, \cite{d0}, and CDF \cite{cdf} collaborations are reproduced
when this color structure of the interactions is superimposed on the usual
perturbative QCD calculation for the production of the hard jets.

Rapidity gaps refer to intervals in pseudo-rapidity devoid of hadronic
activity. The most simple example is the region between the final state
protons, or its excited states, in $pp$ elastic scattering and diffractive
dissociation.  Such processes were first observed in the late 50's in cosmic
rays experiments \cite{c_rays} and have been extensively studied at
accelerators \cite{goul_rep}. Attempts to describe the formation of rapidity
gaps have concentrated on Regge theory and the pomeron \cite{schlein,ua8}, and
on its possible QCD incarnation in the form of a colorless 2-gluon state
\cite{2gluons,sterman}.

After the observation of rapidity gaps in deep inelastic scattering (DIS), it
was suggested \cite{sci} that events with and without rapidity gaps are
identical from a partonic point of view, except for soft color interactions
that, occasionally, lead to a region devoid of color between final state
partons.  We pointed out \cite{us-1} that this soft color mechanism is
identical to the color evaporation mechanism \cite{cem} for computing the
production rates of heavy quark pairs produced in color singlet onium states,
like $J/\psi$.  Moreover, we also suggested that the soft color model could
provide a description for the production of rapidity gaps in hadronic
collisions \cite{us-1}.

\begin{figure}
\parbox[c]{3.in}{
	\mbox{\epsfig{file=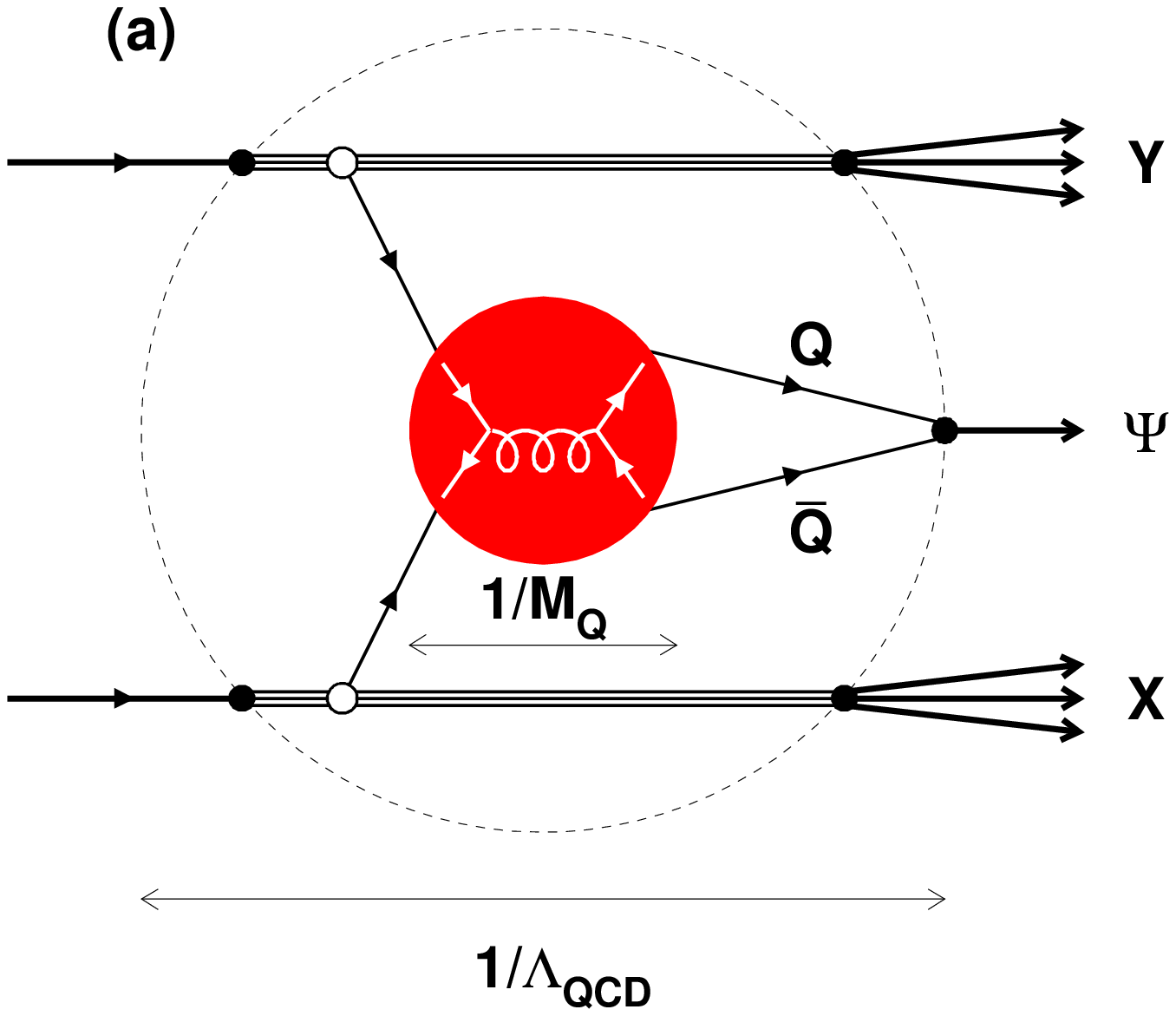,width=\linewidth}}
}
\hfill
\parbox[c]{3.in}{
\mbox{\epsfig{file=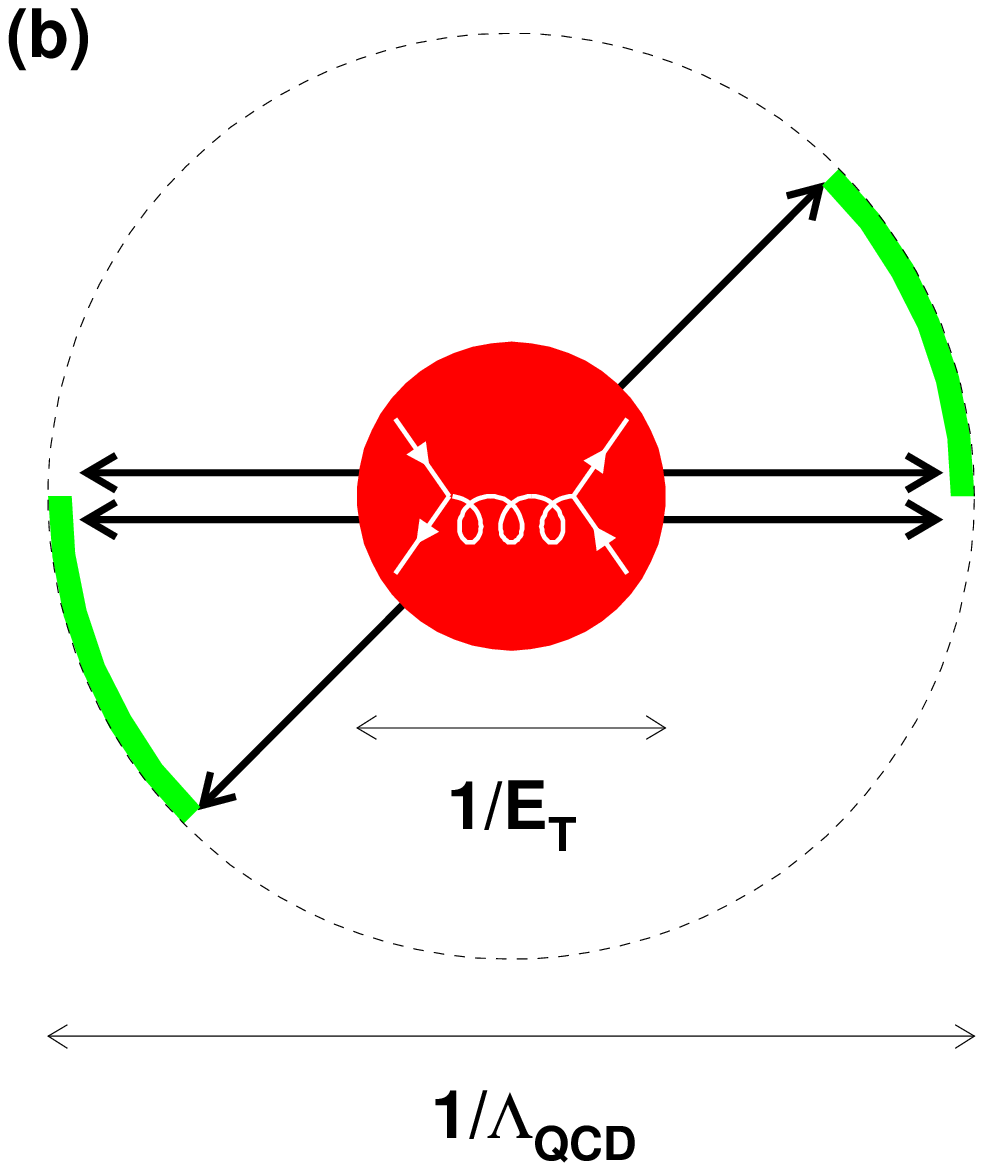,width=0.7\linewidth}}
}
\vskip 12pt
\caption{Sketch of the soft color mechanism for (a) quarkonium production; (b)
rapidity gap formation. We indicate in the figures the typical scale of the
hard scattering and the hadronization scale $1/\Lambda_{QCD}$. The soft color
rearrangement occurs between these two distinct scales.}
\label{fig:quark}
\end{figure}

Color evaporation assumes that quarkonium formation is a two-step process: the
pair of heavy quarks is formed at the perturbative level with scale $M_Q$, and
bound into quarkonium at scale $\Lambda_{QCD}$ (see Fig.\
\ref{fig:quark}a). Heavy quark pairs of any color below open flavor threshold
can form a colorless asymptotic quarkonium state provided they end up in a
color singlet configuration after the inevitable exchange of soft gluons with
the final state spectator hadronic system. The final color state of the quark
pairs is not dictated by the hard QCD process, but by the fate of their color
between the time of formation and emergence as an asymptotic state.

The success of the color evaporation model to explain the data on quarkonium
production is unquestionable \cite{cem-all}.  We show here that the
straightforward application of the color evaporation approach to the string
picture of QCD readily explains the formation of rapidity gaps between jets at
the Tevatron and HERA colliders.

\section{Color Counting Rules}

In the color evaporation scheme for calculating quarkonium production,
it is assumed that all color configurations of the quark pair occur
with equal probability. This must be a reasonable guess because,
before formation as an asymptotic state, the heavy quark pair can
exchange an infinite number of long wavelength soft gluons with the
hadronic final state system in which it is immersed. For instance, the
probability that a $Q\bar{Q}$ pair ends up in a color singlet state is
$1/(1+8)$ because all states in $\bf3 \otimes \bf\bar3 =\bf8
\oplus{\bf1}$ are equally probable.

We propose that the same color counting applies to the final state
partons in high $E_T$ jet production. In complete analogy with
quarkonium, the production of high energy jets is a two-step process
where a pair of high $E_T$ partons is perturbatively produced at a
scale $E_T$, and hadronize into jets at a scale of order
$\Lambda_{QCD}$ by stretching color strings between the partons and
spectators. The strings subsequently hadronize. Rapidity gaps appear
when a cluster of dynamical partons, {\it i.e.} interacting partons or
spectators, form a color singlet (see Fig.\ \ref{fig:quark}b). As
before, the probability for forming a color singlet cluster is
inversely proportional to its color multiplicity.

In this scenario we expect that quark-quark processes possess a higher
probability to form rapidity gaps than gluon-gluon reactions, because of their
smaller color multiplicity. This simple idea is at variance with the two-gluon
exchange model for producing gaps, in which $F_{QQ}<F_{GG}$, where
$F_{QQ(GG)}$ is the gap probability of reactions initiated by quark-quark
(gluon-gluon) collisions.  We already confronted these diverging predictions
using the Tevatron data \cite{previous}. We analyzed the gap fraction in $p
\bar p$ collisions in terms of quark-quark, quark-gluon, and gluon-gluon
subprocesses, {\it i.e.}
\begin{equation}
F_{gap} = \sum_{ij}F_{ij}d\sigma_{ij}/d\sigma \;\; , 
\end{equation}
where $i(j)$ is a quark or a gluon and $d\sigma=\sum_{ij}d\sigma_{ij}$.  We
found that $F_{QQ}>F_{GG}$. This somewhat unexpected feature of the data is in
line with the soft color idea.

In order to better understand the soft color idea let us consider the
formation of rapidity gaps between two jets in opposite hemispheres,
which happens when the interacting parton forming the jet and the
accompanying remnant system form a color singlet. This may occur for
more than one subprocess $N$ and, therefore, the gap fraction is
\begin{equation}
F_{gap} = \frac{1}{d\sigma}~\sum_{N} F_{N} d\sigma_{N}  \;\; ,
\end{equation} 
where $F_N$ is the probability for gap formation in the $N^{th}$
subprocess, $d\sigma_N$ is the corresponding differential
parton-parton cross section, and $d\sigma$ ($ = \sum_N d\sigma_{N}$)
is the total cross section. In our model, the probabilities $F_N$ are
determined by the color multiplicity of the state and spatial
distribution of partons while $d\sigma_N$ is evaluated using
perturbative QCD.

The soft color procedure is obvious in a specific example: let us
calculate the gap formation probability for the subprocesses
\[
p\;\bar{p}\; \to Q^V\bar{Q}^V \to Q\,\bar{Q}\,X\,Y \;\;,
\]
where $Q^V$ stands for $u$ or $d$ valence quark, and $X$ $(Y)$ is the diquark
remnant of the proton (antiproton). The final state is composed of the $X$
($\bf3\otimes\bf3$) color spectator system with rapidity $\eta_X=+\infty$, the
$Y$ ($\bf\bar3\otimes\bf{\bar3}$) color spectator system with
$\eta_Y=-\infty$, one $\bf3$ parton $j_1$, and one $\bf{\bar3}$ parton
$j_2$. It is the basic assumption of the soft color scheme that by the time
these systems hadronize, any color state is equally likely. One can form a
color singlet final state between $X$ and $j_1$ since ${\bf3 \otimes 3 \otimes
3}={\bf 10 \oplus8 \oplus8\oplus1}$, with probability $1/27$. Because of
overall color conservation, once the system $X\otimes j_1$ is in a color
singlet, so is the system $Y\otimes j_2$.  On the other hand, it is not
possible to form a color singlet system with $j_1$ and $Y$.  Moreover, to form
a rapidity gap these systems ($j_1 \otimes X$ and $j_2 \otimes Y$) must not
overlap in rapidity space. Since the experimental data consists of events
where the two jets are in opposite hemispheres, the only additional
requirements are $j_1$ to be in the same hemisphere as $X$, {\it i.e.}
$\eta_1>0$, and $j_2$ to be in the opposite hemisphere ($\eta_1 \cdot \eta_2
<0$). In this configuration, the color strings linking the remnant and the
parton in the same hemisphere will not hadronize in the region between the two
jets.  We have thus produced two jets separated by a rapidity gap using the
color counting rules which form the basis of the color evaporation scheme for
calculating quarkonium production.

As it is clear from the above example, the application of the soft
color model for rapidity gap formation requires the analyses of the
color multiplicity of the possible partonic subprocesses. In the next
sections, we apply this model to the production of rapidity gaps
between jets in photoproduction at HERA and hadronic collisions at the
Tevatron, spelling out the relevant counting rules.

\section{Rapidity Gaps at HERA}

\begin{figure}
\parbox[b]{3.in}{
\mbox{\epsfig{file=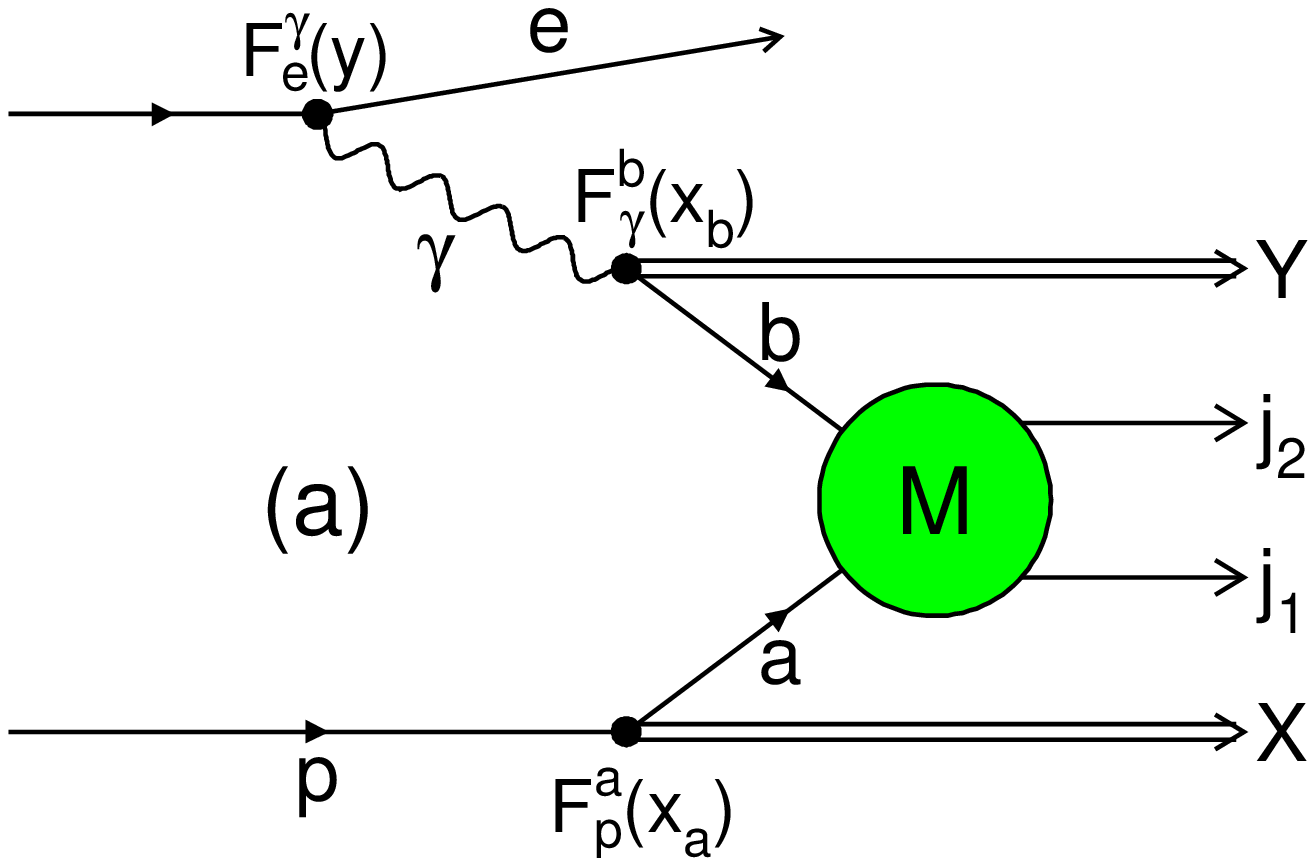,width=\linewidth}}
}
\hfill
\parbox[b]{3.in}{
 \mbox{\epsfig{file=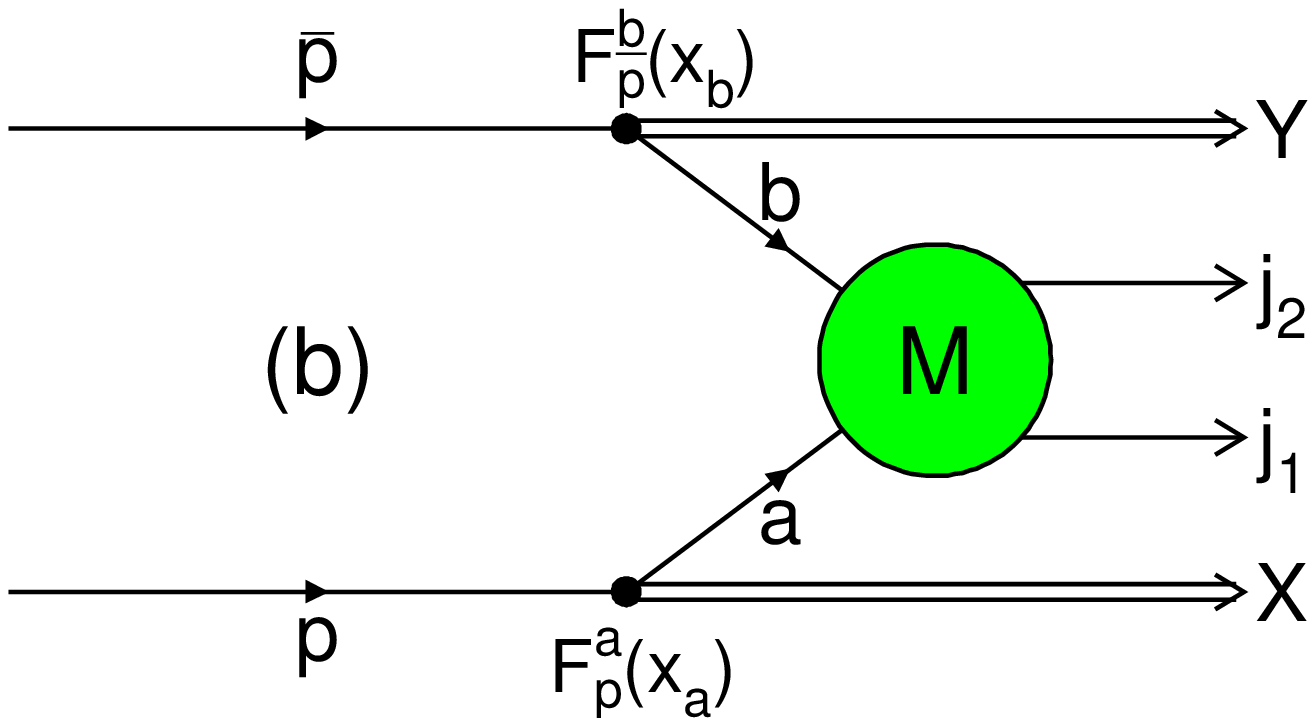,width=\linewidth}}
}
\vskip 12pt
\caption{Kinematics for dijet photoproduction (a) and hadroproduction (b). The
proton and its remnants system $X$ move in the positive rapidity direction.  }
\label{fig:dis}
\end{figure}

The parton diagram for dijet photoproduction is shown in Fig.\  
\ref{fig:dis}a. It is related to the $ep$ cross section by
\begin{equation}
 \sigma_{e p\rightarrow j_1 j_2 X Y}(s)=
\int_{y_{min}}^{y_{max}} \int_{Q^2_{min}}^{Q^2_{max}}
F_e^\gamma (y,Q^2)\, \sigma_{p \gamma\rightarrow j_1 j_2 X Y }(W)
\,dy\,dQ^2 \; \; ,
\end{equation}
where $W$ is the center-of-mass energy of the $p \gamma $ system, $y= W^2/s$
is the fraction of the electron momentum carried by the photon, and $Q^2$ is
the photon virtuality. $Q^2$ ranges from $Q^2_{min} = M_e^2y^2 /(1-y)$ to
$Q^2_{max}$ which depends on the kinematic coverage of the experimental
apparatus. The distribution function of photons in the electron is
\begin{equation}
F_e^\gamma (y,Q^2) =
\frac{\alpha}{2\pi y\,Q^2}\left[1+(1-y)^2-\frac{2M_e^2 y^2}{Q^2}\right]
\;\; ,
\end{equation}
where $M_e$ is the electron mass and $\alpha$ is the fine-structure
constant.

The $p \gamma $ cross section is related to the parton-parton cross
section by
\begin{equation}
\sigma_{p \gamma \rightarrow j_1 j_2 X Y }(W)=
\sum_{a,b}\int\!\!\int F_p^a(x_a)\,F_\gamma^b(x_b)\,
\sigma_{ab\to p_1 p_2}(\hat s)\,dx_a\,dx_b \;\; ,
\end{equation}
where $F_p^a(x_a)$ ($F_\gamma^b(x_b)$) is the distribution function
for parton $a$ ($b$) in the proton (photon) and $\sqrt{\hat s} =
\sqrt{x_a x_b}\,W$ is the parton-parton center-of-mass energy. For
direct $p\gamma$ reactions ($b \equiv \gamma$), $F_\gamma^\gamma (x_b) =
\delta(1-x_b)$. The hadronic system $X$ $(Y)$ is the proton (photon)
remnant, and $j_{1(2)}$ is the jet which is initiated by the parton
$p_{1(2)}$. The proton is assumed to travel in the positive rapidity
direction, and the $t$-channel momentum squared is defined as
$t=(P_a-P_1)^2$, where $P_a$ is the momentum of the parton $a$, and
$P_1$ is the momentum of the parton $p_1$. The expressions for the
parton-parton invariant amplitudes can be found, for instance, in
reference \cite{collider}.

We present in Table \ref{tab:mult} the irreducible decomposition of active
parton systems that yield color singlet states, {\it e. g.}  ${\bf 3 \otimes 8
= 15 \oplus6\oplus3}$ is omitted.  Taking into account this table, it is
simple to obtain the $SU(3)_{color}$ representations and the gap formation
probability for all possible subprocesses.  These are displayed in Table
\ref{tab:gp}.  Notice that only resolved photon processes can produce
rapidity gaps because there is no hadronic remnant associated with direct
photons.

One of the features of the color configurations shown in Table \ref{tab:gp} is
that, for all classes of subprocesses, when a color singlet is (not) allowed
in one of the clusters, the same happens for the other one. Moreover, it can
happen that the color multiplicities are different in the two clusters. In
this case the probability for gap formation is given by the largest of the two
probabilities because, once that cluster forms a color singlet, the other
cluster must do so as well by overall color conservation.

\subsection{ZEUS Results}

The ZEUS collaboration \cite{zeus} has measured the formation of rapidity gaps
between jets produced in $ep$ collisions with $0.2<y<0.85$ and photon
virtuality $Q^2<4$ GeV$^2$. Jets were defined by a cone radius of 1.0 in the
$(\eta,\phi)$ plane, where $\eta$ is the pseudorapidity and $\phi$ is the
azimuthal angle. In the event selection, jets were required to have $E_T>6$
GeV, to not overlap in rapidity ($\Delta\eta= |\eta_1 - \eta_2|>2$), to have a
mean position $|\bar\eta|<0.75$, and to be in the region $\eta<2.5$. The cross
sections were measured in $\Delta\eta$ bins in the range $2 \leq \Delta\eta
\leq 4$.

\begin{figure}
\parbox[c]{3.in}{
\mbox{\epsfig{file=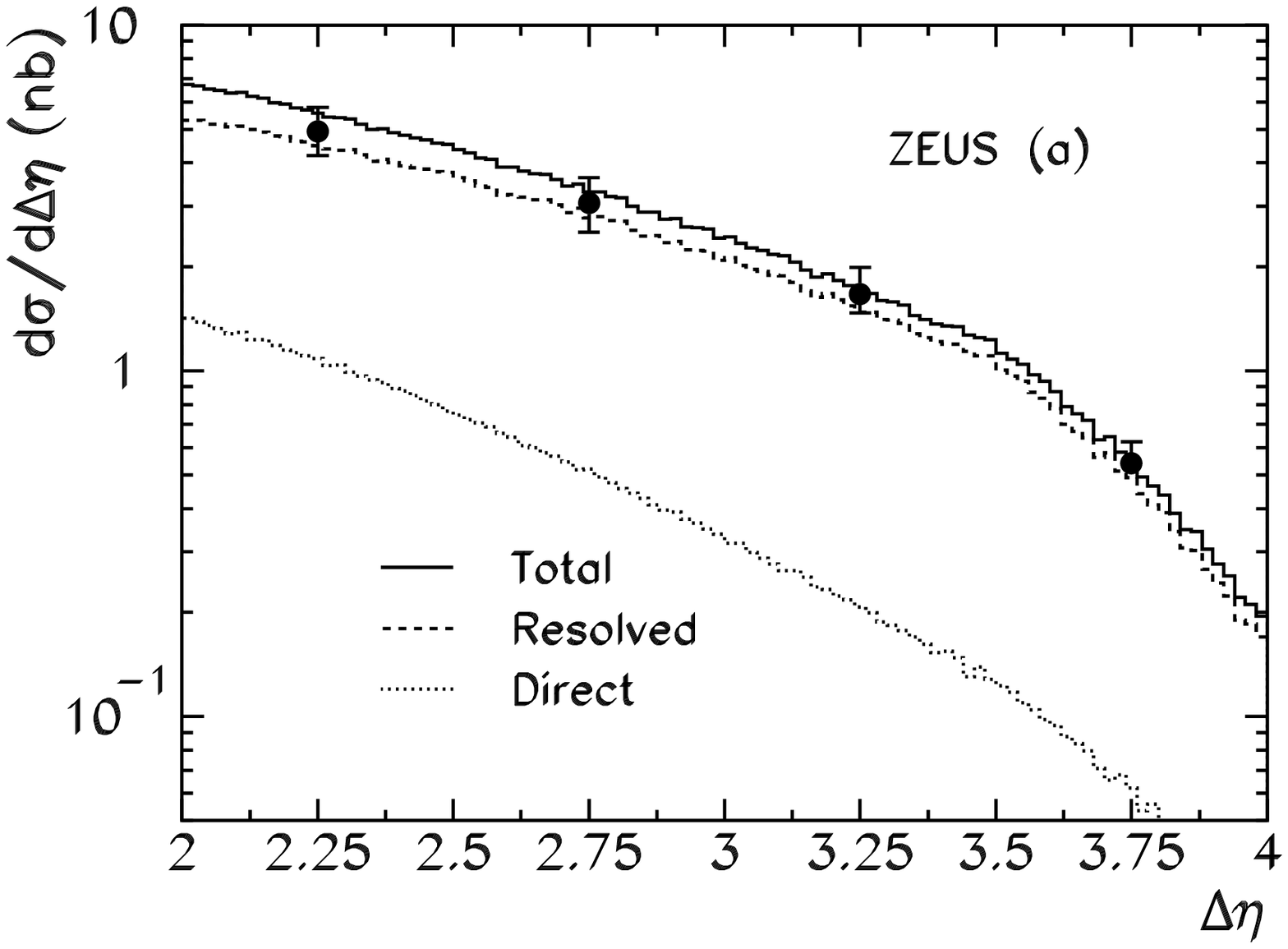,width=\linewidth}}
}
\hfill
\parbox[c]{3.in}
{
\mbox{\epsfig{file=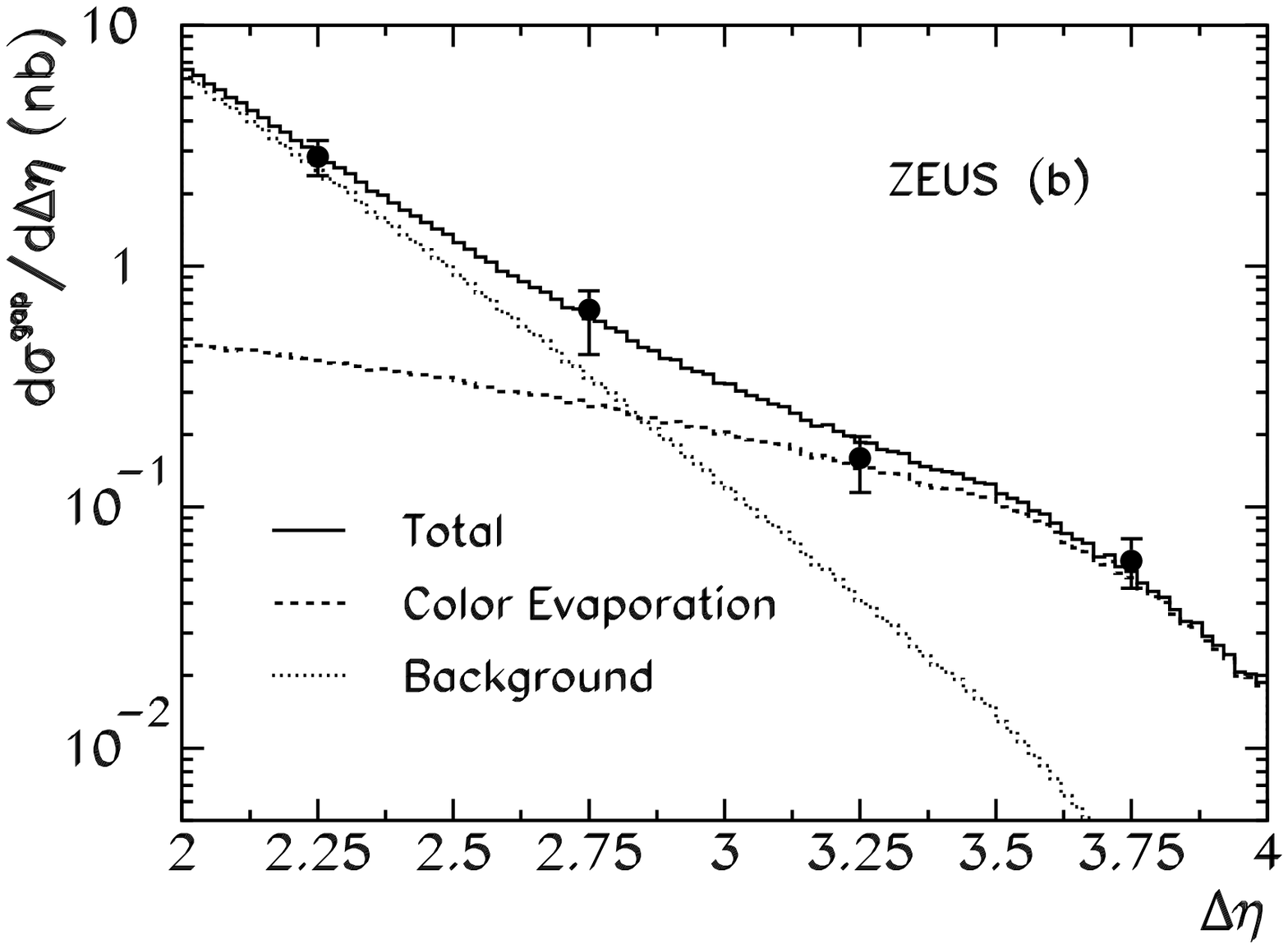,width=\linewidth}} 
}
\vskip 12pt
\caption{Differential dijet cross section as function of the rapidity between 
the jets: (a) all events; (b) events presenting a rapidity gap.  The
points with error bars represent the data obtained by the ZEUS
collaboration \protect\cite{zeus}.}
\label{fig:zeus_dsigma}
\end{figure}

For the above event selection, we evaluated the dijet differential cross
section $d\sigma^{jets}/d\Delta\eta$, which is the sum of the direct
($d\sigma_{dir}$) and the resolved photon ($d\sigma_{res}$) cross sections.
We used the GRV-LO \cite{grv-lo} distribution function for the proton, and the
GRV \cite{grv} for the photon. We fixed the renormalization and factorization
scales at $\mu_R=\mu_F=E_T/2$, and calculated the strong coupling constant for
four active flavors with $\Lambda_{QCD}=350$ MeV. Our results are confronted
with the experimental data in Fig.\ \ref{fig:zeus_dsigma}a, showing that we
describe well both the shape and absolute normalization of the total dijet
cross section. Notice that the bulk of the cross section originates from
resolved events.

Now we turn to dijet events showing a rapidity gap. We evaluate the
differential cross section $d\sigma^{gap}/d\Delta\eta$ which has two sources
of gap events: color evaporation gaps ($d\sigma^{gap}_{cem}$) and background
gaps ($d\sigma^{gap}_{bg}$). In our model, the gap cross section is the
weighted sum over resolved events
\begin{equation}
d\sigma^{gap}_{cem}=\sum_N F_N~ d\sigma^N_{res} \;\; ,
\end{equation}
with the gap probability $F_N$ for the different processes given in Table
\ref{tab:gp}. Background gaps are formed when the region of rapidity between
the jets is devoid of hadrons because of statistical fluctuation of ordinary
soft particle production. Their rate should fall exponentially as the rapidity
separation $\Delta\eta$ between the jets increases \cite{zeus}. We parametrize
the background gap probability as
\begin{equation}
F_{bg}(\Delta\eta) = e^{b(2-\Delta\eta)} \;\; ,
\end{equation}
where $b$ is a constant. The background gap cross section is then written as
\begin{equation}
d\sigma_{bg}^{gap} = F_{bg}(\Delta\eta)~
(d\sigma^{jets} - d\sigma^{gap}_{cem}) \;\; . 
\end{equation}
Notice that the jet definition used by ZEUS implies that the gap cross
section must be equal to the total dijet cross section at $\Delta \eta
= 2 $. This parametrization of the background does take this fact into
account. Moreover, background gaps can be formed in both resolved and
direct processes.

Our results are compared with the experimental data in
Fig.~\ref{fig:zeus_dsigma}b, where we fitted $b=2.9$ and used the same QCD
parameters of Fig.\ \ref{fig:zeus_dsigma}a.  This value of $b$ agrees with
$b=2.7\pm0.3$ found by ZEUS collaboration, when they approximated the
non-background gap fraction by a constant. As we can see from this figure, the
color evaporation model describes very well the gap formation between jets at
HERA.  It is noteworthy that for large values of $\Delta\eta$ the contribution
of the background gap is negligible. In this region the data is correctly
predicted by the color evaporation mechanism alone, with the probability of
gap formation uniquely determined by statistical counting of color states.

\begin{figure}
\parbox[c]{3.in}{
\mbox{\epsfig{file=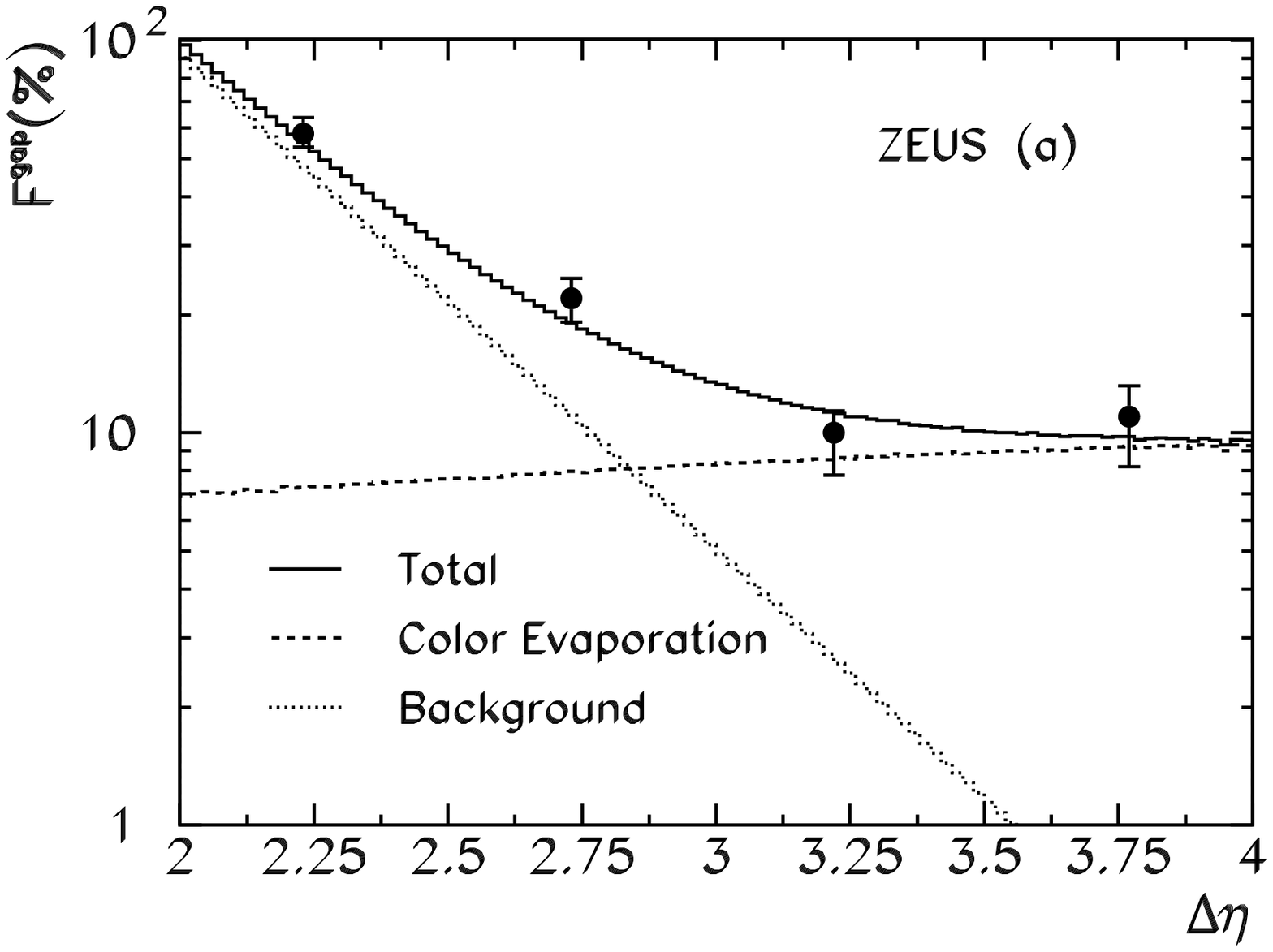,width=\linewidth}}
}
\hfill
\parbox[c]{3.in}{
\mbox{\epsfig{file=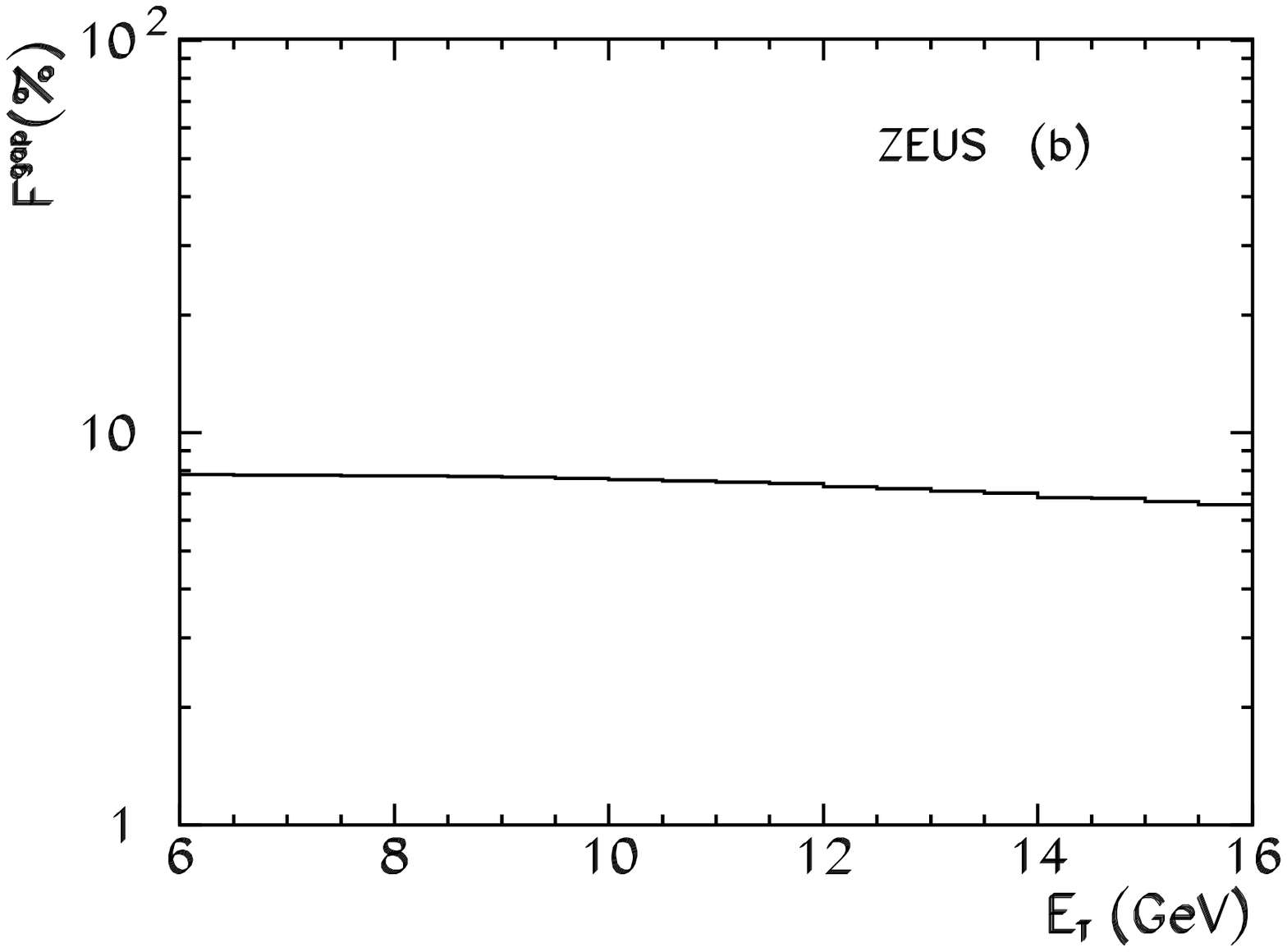,width=\linewidth}}
}
\vskip 12pt
\caption{Fraction of rapidity gap events as a function of: (a) the gap size
$\Delta \eta$; (b) the jet transverse energy $E_T$ at large rapidity
separations ($\Delta \eta >3$). We imposed the cuts used by the ZEUS
collaboration.}
\label{fig:zeus_deta}
\end{figure}

The gap frequency $F^{gap}(\Delta\eta)=d\sigma^{gap}/d\sigma^{jets}$ is shown
in Fig.~\ref{fig:zeus_deta}a, where we show the contributions of the color
evaporation mechanism and the background. Within the color evaporation
framework we can easily predict other differential distributions for the gap
events, which can be used to further test our model.  As an example, we
present in Fig.~\ref{fig:zeus_deta}b the gap frequency predicted by the color
evaporation model as a function of the jet transverse energy for large
rapidity separations ($\Delta \eta >3$), assuming that the background has been
subtracted.  There is currently no data on this distribution.

\subsection{H1 Results}

\begin{figure}
\parbox[c]{3.in}{
\mbox{\epsfig{file=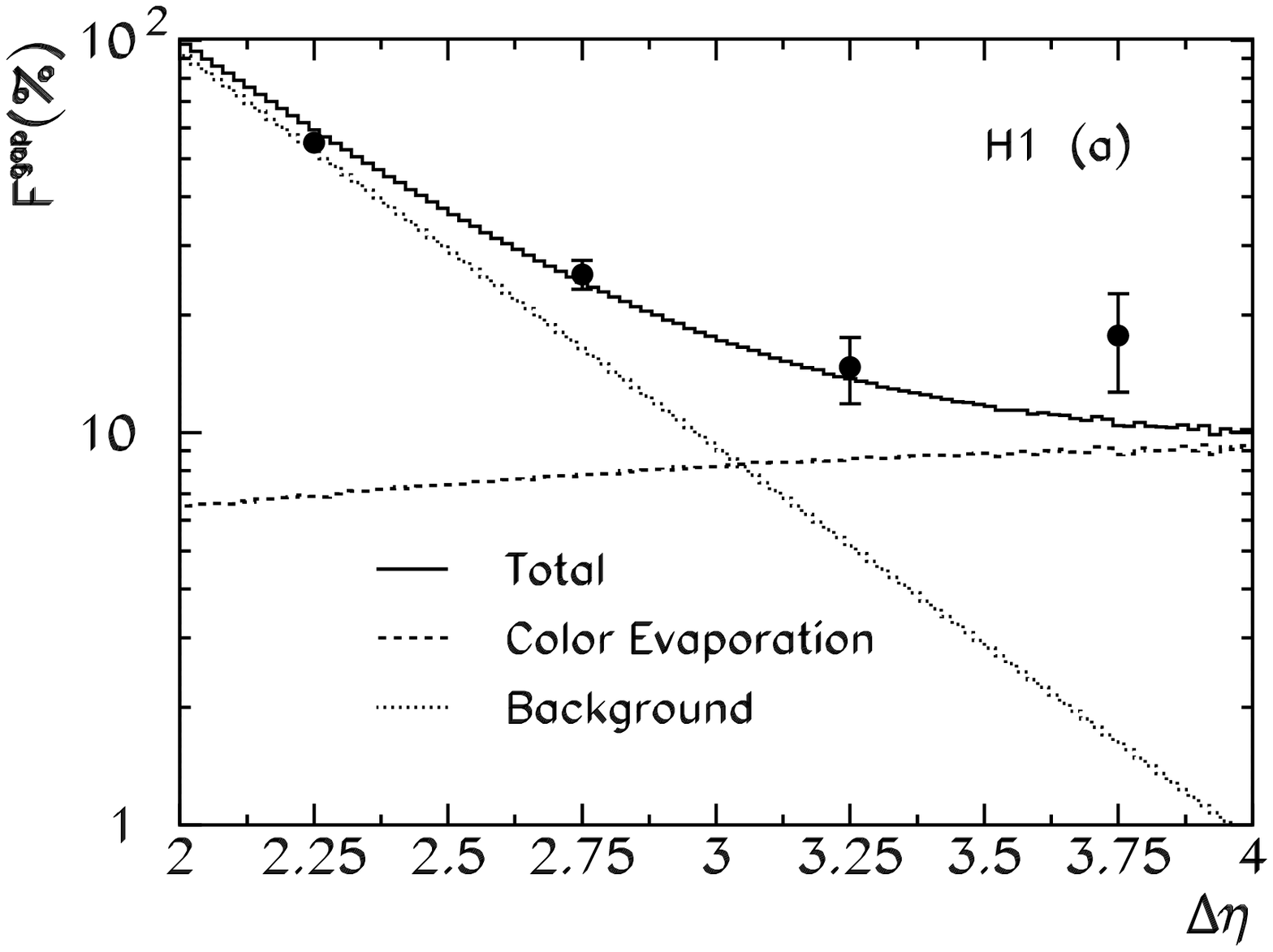,width=\linewidth}}
}
\hfill
\parbox[c]{3.in}{
\mbox{\epsfig{file=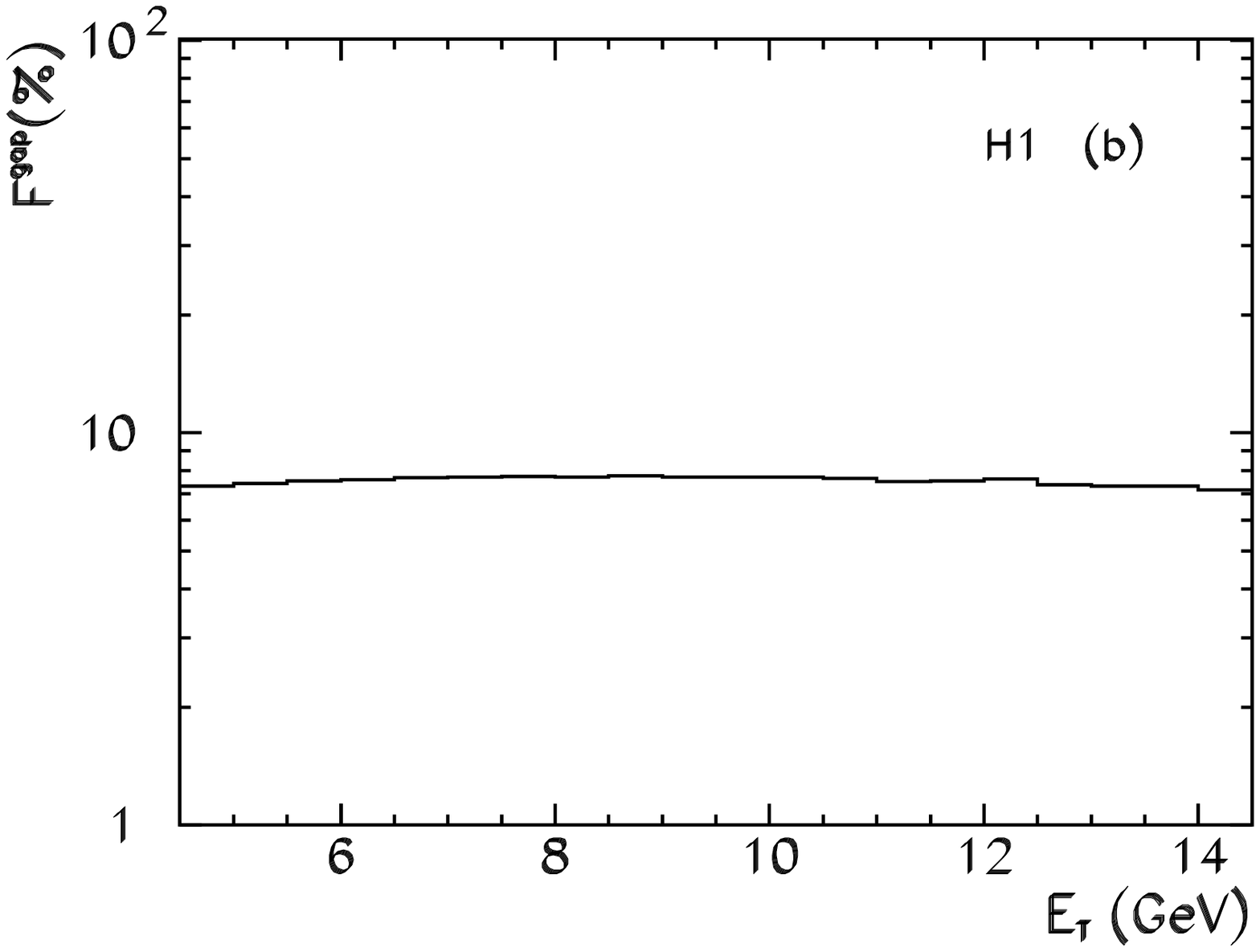,width=\linewidth}}
}
\vskip 12pt
\caption{ Same as in Fig.~\protect\ref{fig:zeus_deta} for the H1 experiment.}
\label{fig:h1_deta}
\end{figure}

We also performed an identical analysis for the data obtained by the H1
collaboration \cite{h1}.  They used the same cone size for the jet definition
$(\Delta R=1)$, and collected events produced in proton-photon reactions with
center-of-mass energy in the range $158<W<247$ GeV and with photon virtuality
$Q^2<0.01$ GeV$^2$. They also imposed cuts on the jets: $-2.82<\eta<2.35$ and
$E_T>4.5$ GeV. Our results are compared with the preliminary experimental data
on Fig.~\ref{fig:h1_deta}a where we used $b=2.3$ to describe the background in
the H1 kinematic range. As before, color evaporation induces gap formation
with a rate compatible with observation. We show in Fig.~\ref{fig:h1_deta}b
our predictions for the background subtracted gap frequency as a function of
the jet transverse energy for large rapidity separations $\Delta \eta > 3$.

\subsection{Survival Probability at HERA}

Our computation of gap rates using color evaporation is free of parameters and
therefore predicts absolute rates, as well as their dependence on kinematic
variables. In practice, this prediction is diffused by the necessity to
introduce a gap survival probability $S_p$, which accounts for the fact that
genuine gap events, as predicted by the theory, can escape experimental
identification because additional partonic interactions in the same event
produce secondaries which spoil the gap. Its value has been estimated for high
energy $p \bar p$ interactions to be of order a few tens of percent. The fact
that the color evaporation calculation correctly accommodates the absolute gap
rate observed in $p\gamma$ collisions implies that $S_p=1$. There is a simple
explanation for this value.  The dijet cross section is dominated by resolved
photons.  However, for resolved processes, a secondary partonic interaction
which could fill the gap is unlikely because it requires resolving the photon
in 2 partons. Although this routinely happens at high energies for hadrons, it
does not for photons.

\section{Rapidity Gaps at Tevatron}

The kinematics for dijet production in $p\bar{p}$ collisions is
illustrated in Fig.~\ref{fig:dis}b, where we denoted by $X$ ($Y$) the
proton (antiproton) remnant, and $j_{1(2)}$ is a parton giving rise to
a jet. The proton is assumed to travel in the positive rapidity
direction. The dijet production cross section is related to the
parton-parton one via
\begin{equation}
\sigma_{p \bar p \rightarrow j_1 j_2 X Y }(s)=
\sum_{a,b}\int\!\!\int F_p^a(x_a)\,F_{\bar p}^b(x_b)\,
\sigma_{ab\to p_1 p_2}(\hat s)\,dx_a\,dx_b \;\; ,
\end{equation}
where $s$ ($\hat s = x_a x_b s$) is the (subprocess) center-of-mass energy
squared and $F^{a(b)}_{p(\bar p)}$ is the distribution function for the parton
$a$ $(b)$. We evaluated the dijet cross sections using MRS-J distribution
functions \cite{mrs-j} with renormalization and factorization scales
$\mu_R=\mu_F=\sqrt{\hat s}$.

The color evaporation model prediction for the gap production rates in $p\bar
p$ collisions is analogous to the one in $p\gamma$ interactions, with the
obvious replacement of the photon by the antiproton, represented as a ${\bf
\bar3\otimes\bar3\otimes\bar3}$ system.  The color subprocesses and their
respective gap formation probabilities are listed in Table \ref{tab:pp}.

Both experimental collaborations presented their data with the background
subtracted.  The CDF collaboration measured the appearance of rapidity gaps at
two different $p\bar p$ center-of-mass energies.  For the data taken at
$\sqrt{s}=1800$ GeV, they required that both jets to have $E_T>20$ GeV, and to
be produced in opposite sides ($\eta_1\cdot \eta_2<0$) within the region
$1.8<|\eta|<3.5$.  For the lower energy data, $\sqrt{s}=630$ GeV, they
required both jets to have $E_T>8$ GeV, and to be produced in opposite sides
within the region $|\eta|>1.8$. Since the experimental distributions are
normalized to unity, on average, we do not need to introduce an {\em ad-hoc}
gap survival probability. Therefore, our predictions do not exhibit any free
parameter to be adjusted, leading to a important test of the color evaporation
mechanism.

\begin{figure}
\mbox{\qquad\epsfig{file=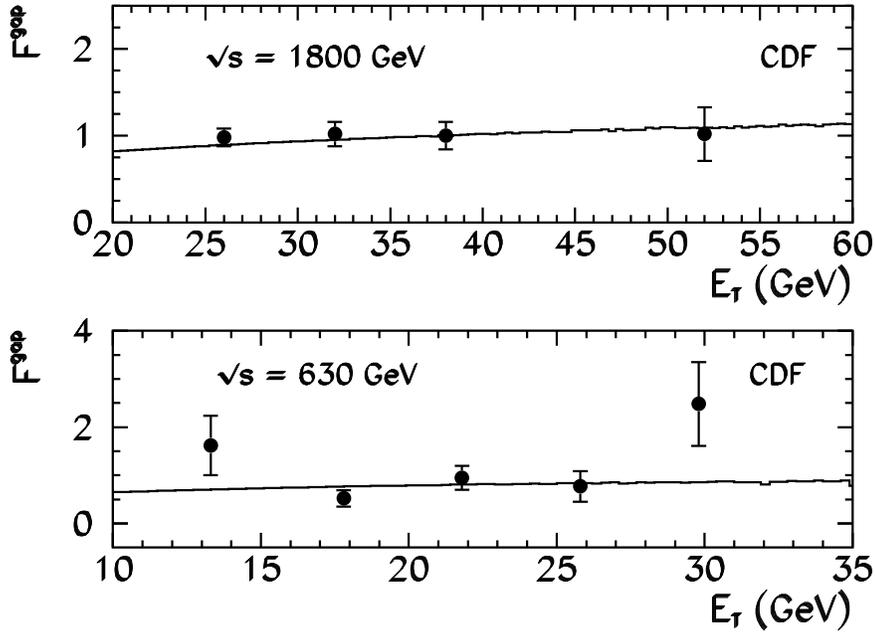,width=0.7\linewidth}}
\vskip 12pt
\caption{Dependence of the gap frequency on the jet transverse energy 
as measured by the CDF collaboration. The absolute normalization 
is arbitrary.}
\label{fig:cdf_et}
\end{figure}

\begin{figure}
\mbox{\qquad\epsfig{file=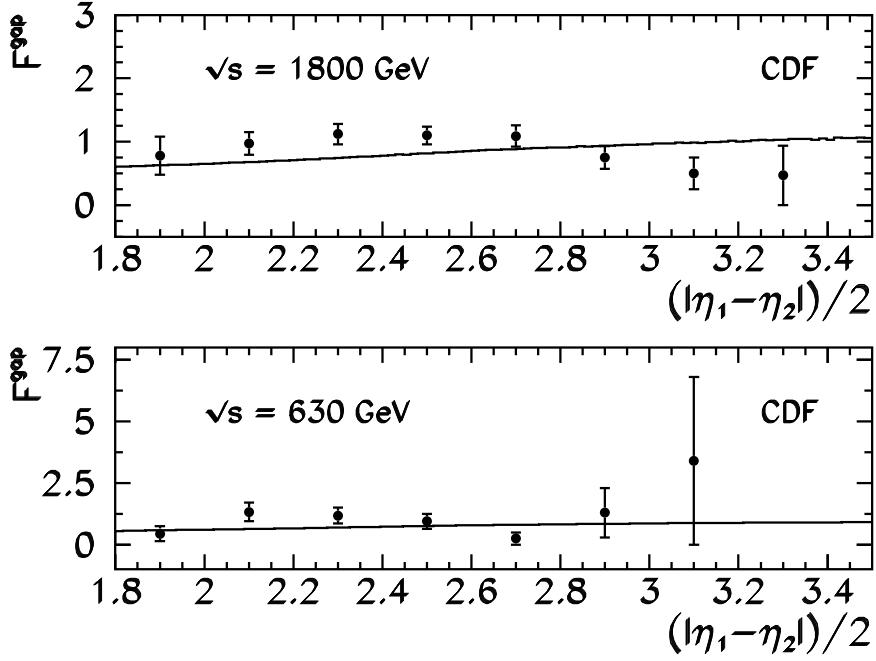,width=0.7\linewidth}}
\caption{The same as Fig.\ \protect\ref{fig:cdf_et} for
half the gap size.}
\label{fig:cdf_deta}
\end{figure}

\begin{figure}
\mbox{\qquad\epsfig{file=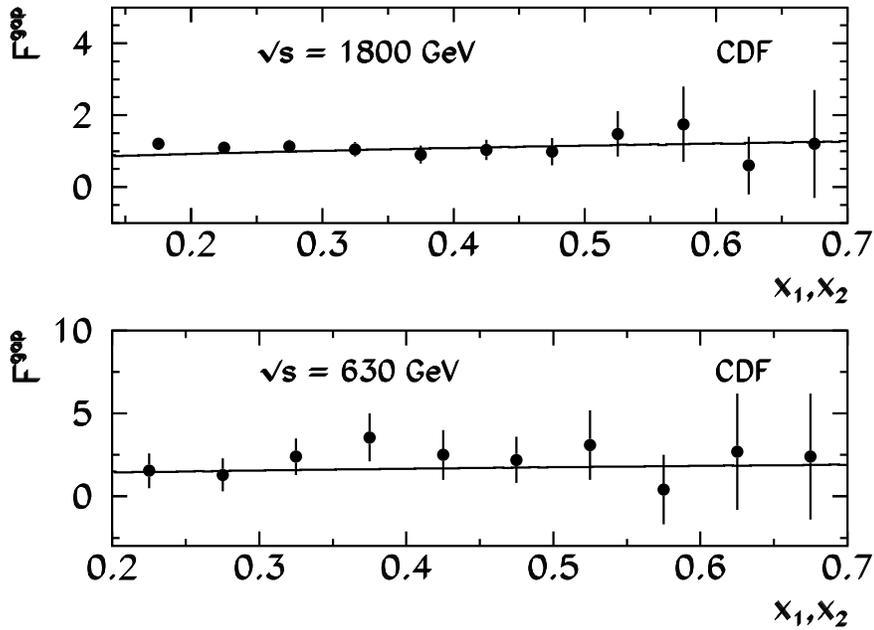,width=0.7\linewidth}}
\vskip 12pt
\caption{The same as Fig.\ \protect\ref{fig:cdf_et} for the
Bjorken-$x$ of each jet. Two entries per event are included in the
distribution.}
\label{fig:cdf_x}
\end{figure}

In Figs.\ \ref{fig:cdf_et}, \ref{fig:cdf_deta}, and \ref{fig:cdf_x} we compare
our predictions with the experimental observations of the gap fraction as a
function of the jets transverse energy, their separation in rapidity, and the
Bjorken-$x$ of the colliding partons, respectively. As we can see, the overall
performance of the color evaporation model is good since it describes
correctly the shape of almost all distributions. This is an impressive result
since the model has no free parameters to be adjusted.

\begin{figure}
\mbox{\qquad\epsfig{file=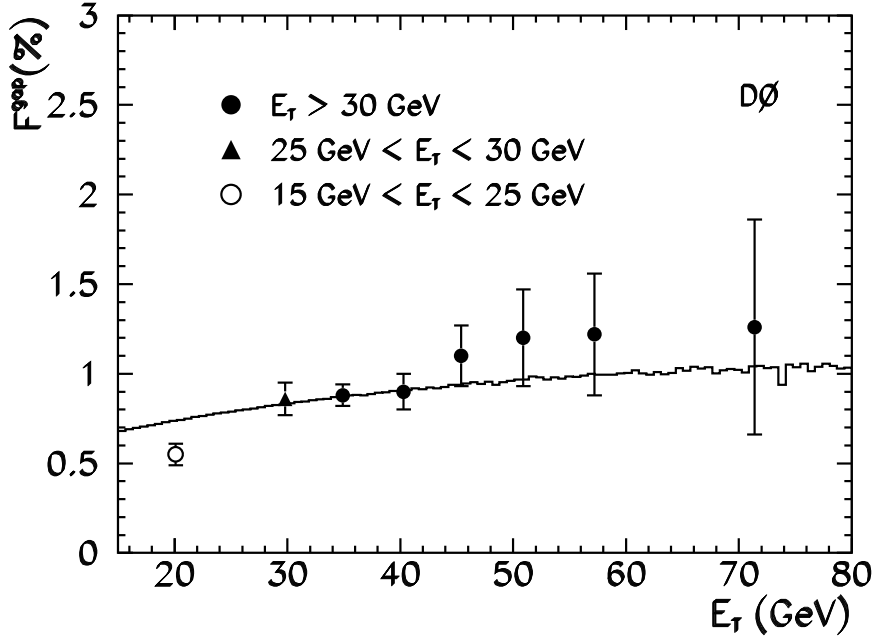,width=0.7\linewidth}}
\vskip 12pt
\caption{Gap fraction as a function of the jet $E_T$ as measured by the D\O\
collaboration at $\protect\sqrt{s}=1800$ GeV. We used a
gap survival probability of $30\%$ to obtain the normalization shown.}
\label{fig:d0_et}
\end{figure}

\begin{figure}
\mbox{\qquad\epsfig{file=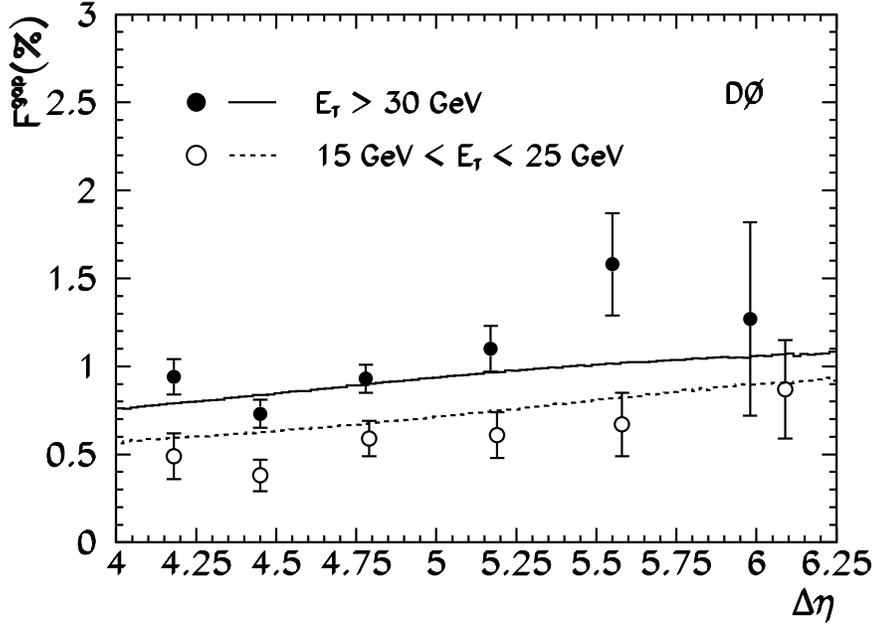,width=0.7\linewidth}}
\vskip 12pt
\caption{Gap fraction as a function of the jets rapidity separation for
two different values of the jets transverse energy. The survival
probability is the same as in Fig.~\protect\ref{fig:d0_et}.}
\label{fig:d0_deta}
\end{figure}

\begin{figure}
\mbox{\qquad\epsfig{file=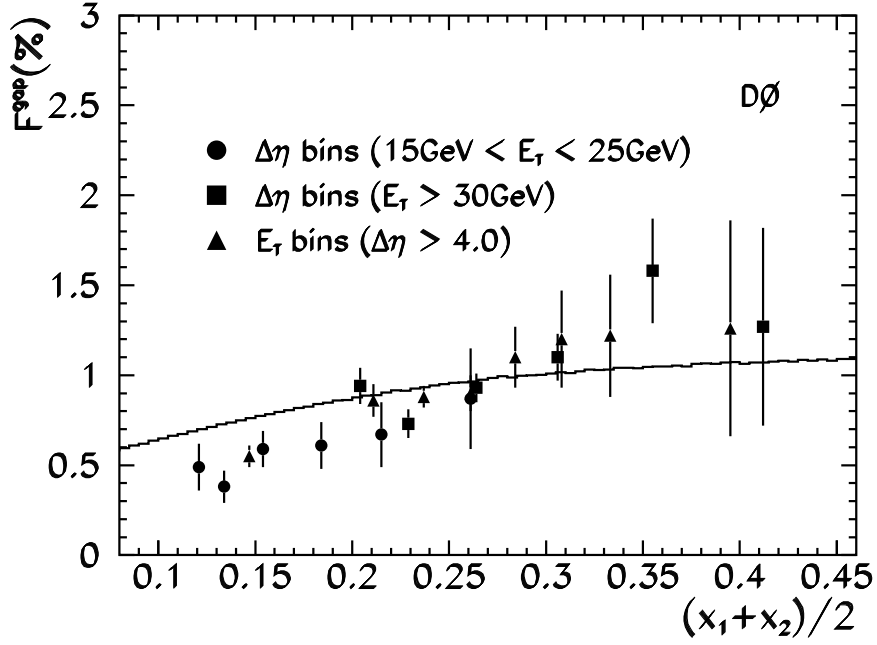,width=0.7\linewidth}}
\vskip 12pt
\caption{Gap fraction as a function of the average Bjorken-$x$ of the two
jets for events collected in different rapidity intervals
and jet transverse energy bins. The survival
probability is the same as in Fig.~\protect\ref{fig:d0_et}.}
\label{fig:d0_xb}
\end{figure}

The D\O\ collaboration has made similar observations at $\sqrt{s}=1800$ GeV.
They required that both jets to have $E_T>15$ GeV, to be produced in opposite
sides ($\eta_1 \cdot \eta_2<0$) within the region $1.9<|\eta|<4.1$, and to be
separated by $|\Delta\eta|>4.0$. In Fig.\ \ref{fig:d0_et} our results are
compared with experimental observations of the dependence of the gap frequency
on jet transverse energy, where we used a gap survival probability $S_p=30\%$
to reproduce the absolute normalization. This is consistent with qualitative
theoretical estimates; see discussion below.  As we can see, the fraction of
gap events increases with the transverse energy of the jets. This is expected
once the dominant process for the rapidity gap formation is quark-quark
fusion, which becomes more important at larger $E_T$.  Apart from the lowest
transverse energy bin, data and theory are in good agreement.  In Fig.\
\ref{fig:d0_deta} we compare our prediction for the dependence of the gap
frequency with the separation between the jets.  Agreement is satisfactory
although the absolute value of our predictions for low transverse energy is
somewhat higher than data as shown in Fig.\ \ref{fig:d0_et}. Finally, in Fig.\
\ref{fig:d0_xb} we show our results for the mean value of the Bjorken-$x$ of
the events, where all correlations between the jet transverse energy and
rapidity have been included. Again, the agreement between theory and data is
satisfactory except for the low transverse energy bins.

\subsection{Survival Probability at Tevatron}

We estimated the survival probability of rapidity gaps formed at $p\bar p$
collisions, comparing our predictions with the values of gap fraction actually
observed. Assuming that the survival probability varies only with the
collision center-of-mass energy, and not with the jet's transverse energy, we
evaluated the average survival probability
\begin{equation}
	S_p= \frac{F^{gap}_{exp}}{F^{gap}_{cem} } \;\; . 
\end{equation}
In order to extract $\bar{S}_p$ we combined the D\O\ and CDF available
data at each center-of-mass energy: 630 and 1800 GeV.  We found
$\bar{S}_p(1800) = 34.4 \pm 3.3$\% and $\bar{S}_p(630) = 65.4 \pm
12$\%, a value compatible with the calculation of Ref.\ \cite{surv}
based on the Regge model, which yields $S_p(1800) = 32.6$\%. For
individual contributions and further details see Table
\ref{tab:surv}. Moreover, we have that $\bar{S}_p(630)/\bar{S}_p(1800)
= 1.9 \pm 0.4$, which is compatible with the theoretical expectation
$2.2 \pm 0.2$ obtained in Ref.\ \cite{sfrac}.

Using the extracted values of the survival probability, we contrasted the color
evaporation model predictions for the gap fraction corrected by $\bar{S}_p$
($F^{gap}_{cor} = F^{gap}_{cem} \times \bar{S}_p$) with the experimental data
in Table \ref{tab:surv}.  We can also compare the ratio $R = F^{gap}_{cor}
(630)/F^{gap}_{cor}(1800)$ with the experimental result.  D\O\ has measured
this fraction for jets with $E_T>12$ GeV for both energies, and they found
$R=3.4\pm1.2$; we predict $R=2.5\pm0.5$. On the other hand, CDF measured this
ratio using different values for $E_T^{min}$ at 630 GeV and 1800 GeV; they
obtained $R=2.0\pm0.9$ while we obtained $R=2.0\pm0.4$ for the same
kinematical arrangement.

\section{Conclusion}

In summary, the occurrence of rapidity gaps between hard jets can be understood
by simply applying the soft color, or color evaporation, scheme for
calculating quarkonium production, to the conventional perturbative QCD
calculation of the production of hard jets.  The agreement between data and
this model is impressive.

\acknowledgments

This research was supported in part by the University of Wisconsin Research
Committee with funds granted by the Wisconsin Alumni Research Foundation, by
the U.S.\ Department of Energy under grant DE-FG02-95ER40896, by Conselho
Nacional de Desenvolvimento Cient\'{\i}fico e Tecnol\'ogico (CNPq), by
Funda\c{c}\~ao de Amparo \`a Pesquisa do Estado de S\~ao Paulo (FAPESP), and
by Programa de Apoio a N\'ucleos de Excel\^encia (PRONEX).


\newpage


\begin{table}
\begin{tabular}{lc}
Final state color multiplicity & Color singlet fraction
\\ \hline
${\bf3}\otimes{\bf\bar3}={\bf8}\oplus{\bf1}$
& 1/9
\\
${\bf3}\otimes{\bf3}\otimes{\bf3}={\bf10}\oplus2({\bf8})\oplus{\bf1}$
& 1/27
\\
${\bf3}\otimes{\bf\bar3}\otimes{\bf8}
={\bf27}\oplus2({\bf10})\oplus3({\bf8})\oplus{\bf1}$
& 1/72
\\
${\bf3}\otimes{\bf3}\otimes{\bf3}\otimes{\bf8}
={\bf35}\oplus3({\bf27})\oplus5({\bf10})
\oplus6({\bf8})\oplus2({\bf1})$
& 2/216
\\
${\bf3}\otimes{\bf3}\otimes{\bf3}\otimes{\bf3}\otimes{\bf\bar3}
={\bf35}\oplus3({\bf27})\oplus6({\bf10})
\oplus8({\bf8})\oplus3({\bf1})$
& 3/243
\end{tabular}
\vskip 12pt
\caption[]
{Irreducible decomposition of relevant $SU(3)$ representations. Only
those that generate singlets are shown.}
\label{tab:mult}
\end{table}

\begin{table}
\begin{tabular}{lccccc}
Subprocess & $j_1$ & $j_2$ & $X$ & $Y$ & $F_N$
\\ \hline
$Q^V Q \rightarrow QQ$ &
${\bf 3} $&$ {\bf 3} $&$ {\bf 3\otimes3} $&$ {\bf \bar3 } $
&$1/9\; \forall\; \eta_1$
\\
$Q^S Q \rightarrow QQ$ &
${\bf 3} $&$ {\bf 3} $&$ {\bf 3\otimes3\otimes3\otimes\bar3} $
&$ {\bf \bar3}  $&$1/9\; \forall\; \eta_1$
\\
$\bar Q^S \bar Q \rightarrow \bar Q \bar Q$ &
${\bf \bar3} $&$ {\bf \bar3} $&$ {\bf 3\otimes3\otimes3\otimes3} $
&$ {\bf 3}  $&$1/9\; \forall\; \eta_1$
\\
$Q^V \bar Q \rightarrow Q\bar Q$ &
${\bf 3} $&$ {\bf \bar3} $&$ {\bf 3\otimes3} $&$ {\bf 3}  $
&$1/9 $ for $\eta_1>\eta_2$
\\
$Q^S \bar Q \rightarrow Q\bar Q$ &
${\bf 3} $&$ {\bf \bar3} $&$ {\bf 3\otimes3\otimes3\otimes\bar3} $
&$ {\bf 3}  $&$1/9 $ for $\eta_1>\eta_2$
\\
$\bar Q^S Q \rightarrow \bar Q Q$ &
${\bf \bar3} $&$ {\bf 3} $&$ {\bf 3\otimes3\otimes3\otimes3} $
&$ {\bf \bar3}  $&$1/9 $ for $\eta_1>\eta_2$
\\
$Q^V \bar Q \rightarrow GG$ &
${\bf 8} $&$ {\bf 8} $&$ {\bf 3\otimes3} $&$ {\bf 3 }  $
&$0 $
\\
$Q^S \bar Q \rightarrow GG$ &
${\bf 8} $&$ {\bf 8} $&$ {\bf 3\otimes3\otimes3\otimes\bar3}  $
&$ {\bf 3}  $&$0 $
\\
$\bar Q^S  Q \rightarrow GG$ &
${\bf 8} $&$ {\bf 8} $&$ {\bf 3\otimes3\otimes3\otimes3}  $
&$ {\bf \bar3}  $&$0 $
\\
$Q^V G \rightarrow QG$ &
${\bf 3} $&$ {\bf 8} $&$ {\bf 3\otimes3} $&$ {\bf 3\otimes\bar3 }  $
&$1/27$ for $\eta_1>\eta_2$
\\
$Q^S G \rightarrow QG$ &
${\bf 3} $&$ {\bf 8} $&$ {\bf 3\otimes3\otimes3\otimes\bar3}  $
&$ {\bf 3\otimes\bar3 }  $&$1/72 $ for $\eta_1>\eta_2$
\\
$\bar Q^S G \rightarrow \bar QG$ &
${\bf \bar3} $&$ {\bf 8} $&$ {\bf 3\otimes3\otimes3\otimes3}  $
&$ {\bf 3\otimes\bar3 }  $&$1/72 $ for $\eta_1>\eta_2$
\\
$G Q \rightarrow G Q $ &
${\bf 8} $&$ {\bf 3} $&$ {\bf 3\otimes3\otimes3} $&$ {\bf \bar3}   $
&$1/9 $ for $\eta_1>\eta_2$
\\
$G \bar Q \rightarrow G \bar Q $ &
${\bf 8} $&$ {\bf \bar3} $&$ {\bf 3\otimes3\otimes3} $&$ {\bf 3}   $
&$1/9 $ for $\eta_1>\eta_2$
\\
$G G \rightarrow Q\bar Q $ &
${\bf 3} $&$ {\bf \bar3} $&$ {\bf 3\otimes3\otimes3} $
&$ {\bf 3\otimes\bar3 }  $ & $0 $
\\
$G G \rightarrow G G $ &
${\bf 8} $&$ {\bf 8} $&$ {\bf 3\otimes3\otimes3} $
&$ {\bf 3\otimes\bar3 } $ &$1/72\; \forall \; \eta_1$
\end{tabular}
\vskip 12pt
\caption[] {Color multiplicities and gap probabilities $F_N$ for the reaction
$p\,\gamma \to j_1\,j_2\,X\,Y$, where $X$ and $Y$ are respectively the proton
and the photon remnant systems.  $Q^{V(S)}$ stands for valence (sea) quarks,
and we assumed that the proton travels in the positive rapidity direction.}
\label{tab:gp}
\end{table}


\begin{table}
\begin{tabular}{lccccc}
Subprocess & $j_1$ & $j_2$ & $X$ & $Y$ & $F_N$
\\ \hline
$Q^V Q^S \rightarrow QQ$ &
${\bf 3} $&$ {\bf 3} $&$ {\bf 3\otimes3} $
&$ {\bf \bar3\otimes\bar3\otimes\bar3\otimes\bar3} $
&$1/27 \;\forall\;\eta_1$
\\
$Q^S Q^S \rightarrow QQ$ &
${\bf 3} $&$ {\bf 3} $&$ {\bf 3\otimes3\otimes3\otimes\bar3} $
&$ {\bf \bar3\otimes\bar3\otimes\bar3\otimes\bar3} $&$1/81 \;\forall\;\eta_1$
\\
$\bar Q^S \bar Q^V \rightarrow \bar Q\bar Q$ &
${\bf \bar3} $&$ {\bf \bar3} $&$ {\bf 3\otimes3\otimes3\otimes3} $
&$ {\bf \bar3\otimes\bar3}  $&$1/27 \;\forall\;\eta_1$
\\
$\bar Q^S \bar Q^S \rightarrow \bar Q\bar Q$ &
${\bf \bar3} $&$ {\bf \bar3} $&$ {\bf 3\otimes3\otimes3\otimes3} $
&$ {\bf \bar3\otimes\bar3\otimes\bar3\otimes3}  $&$1/81 \;\forall\;\eta_1$
%
\\
$Q^V \bar Q^V \rightarrow Q\bar Q$ &
${\bf 3} $&$ {\bf \bar3} $&$ {\bf 3\otimes3} $&$ {\bf \bar3\otimes\bar3}  $
&$1/27 $ for $\eta_1>0$
\\
$Q^V \bar Q^S \rightarrow Q\bar Q$ &
${\bf 3} $&$ {\bf \bar3} $&$ {\bf 3\otimes3} $
&$ {\bf \bar3\otimes\bar3\otimes\bar3\otimes3} $
&$1/27$ for $\eta_1>0$
\\
$Q^S \bar Q^V \rightarrow Q\bar Q$ &
${\bf 3} $&$ {\bf \bar3} $&$ {\bf 3\otimes3\otimes3\otimes\bar3} $
&$ {\bf \bar3\otimes\bar3}  $&$1/27 $ for $\eta_1>0$
\\
$Q^S \bar Q^S \rightarrow Q\bar Q$ &
${\bf 3} $&$ {\bf \bar3} $&$ {\bf 3\otimes3\otimes3\otimes\bar3} $
&$ {\bf \bar3\otimes\bar3\otimes\bar3\otimes3}  $
&$1/81 $ for $\eta_1>0$
\\
$\bar Q^S Q^S \rightarrow \bar Q Q$ &
${\bf \bar3} $&$ {\bf 3} $&$ {\bf 3\otimes3\otimes3\otimes3} $
&$ {\bf \bar3\otimes\bar3\otimes\bar3\otimes\bar3}  $
&$1/81 $ for $\eta_1>0$
%
\\
$Q^V \bar Q^V \rightarrow GG$ &
${\bf 8} $&$ {\bf 8} $&$ {\bf 3\otimes3} $&$ {\bf \bar3\otimes\bar3}  $
&$0 $
\\
$Q^V \bar Q^S \rightarrow GG$ &
${\bf 8} $&$ {\bf 8} $&$ {\bf 3\otimes3} $
&$ {\bf \bar3\otimes\bar3\otimes\bar3\otimes3}  $&$0 $
\\
$Q^S \bar Q^V \rightarrow GG$ &
${\bf 8} $&$ {\bf 8} $&$ {\bf 3\otimes3\otimes3\otimes\bar3}  $
&$ {\bf \bar3\otimes\bar3}  $&$0 $
\\
$Q^S \bar Q^S \rightarrow GG$ &
${\bf 8} $&$ {\bf 8} $&$ {\bf 3\otimes3\otimes3\otimes\bar3}  $
&$ {\bf \bar3\otimes\bar3\otimes\bar3\otimes3}  $&$0 $
\\
$\bar Q^S Q^S \rightarrow GG$ &
${\bf 8} $&$ {\bf 8} $&$ {\bf 3\otimes3\otimes3\otimes3}  $
&$ {\bf \bar3\otimes\bar3\otimes\bar3\otimes\bar3}  $&$0 $
%
\\
$Q^V G \rightarrow QG$ &
${\bf 3} $&$ {\bf 8} $&$ {\bf 3\otimes3} $
&$ {\bf \bar3\otimes\bar3\otimes\bar3}$ & $1/27$ for $\eta_1>0$
\\
$Q^S G \rightarrow QG$ &
${\bf 3} $&$ {\bf 8} $&$ {\bf 3\otimes3\otimes3\otimes\bar3}  $
&$ {\bf \bar3\otimes\bar3\otimes\bar3 } $ & $1/108 $ for $\eta_1>0$
\\
$\bar Q^S G \rightarrow \bar QG$ &
${\bf \bar 3} $&$ {\bf 8} $&$ {\bf 3\otimes3\otimes3\otimes3}  $
&$ {\bf \bar3\otimes\bar3\otimes\bar3 } $ & $1/108 $ for $\eta_1>0$
\\
$G \bar Q^V \rightarrow G \bar Q $ &
${\bf 8} $&$ {\bf \bar3} $&$ {\bf 3\otimes3\otimes3} $
&$ {\bf \bar3\otimes\bar3 }   $ &$1/27 $ for $\eta_1>0$
\\
$G \bar Q^S \rightarrow G \bar Q $ &
${\bf 8} $&$ {\bf \bar3} $&$ {\bf 3\otimes3\otimes3} $
&$ {\bf \bar3\otimes\bar3\otimes\bar3\otimes3}$
&$1/108 $ for $\eta_1>0$
\\
$G Q^S \rightarrow G Q $ &
${\bf 8} $&$ {\bf 3} $&$ {\bf 3\otimes3\otimes3} $
&$ {\bf \bar3\otimes\bar3\otimes\bar3\otimes\bar3}$
&$1/108$ for $\eta_1>0$
%
\\
$G G \rightarrow Q\bar Q $ &
${\bf 3} $&$ {\bf \bar3} $&$ {\bf 3\otimes3\otimes3} $
&$ {\bf\bar3\otimes\bar3\otimes\bar3 }  $ &$0 $
\\
$G G \rightarrow G G $ &
${\bf 8} $&$ {\bf 8} $&$ {\bf 3\otimes3\otimes3} $
&$ {\bf\bar3\otimes\bar3\otimes\bar3 }  $ & $1/108 \;\forall\;\eta_1 $
\end{tabular}
\vskip 12pt 
\caption[] {Color multiplicities and gap probabilities $F_N$ for the reaction
$p\,\bar p \to j_1\,j_2\,X\,Y$, where $X$ and $Y$ are respectively the proton
and the antiproton remnant systems.  $Q^{V(S)}$ stands for valence (sea)
quarks, and we assumed that the proton travels in the positive rapidity
direction.  }
\label{tab:pp}
\end{table}


\begin{table}
\begin{tabular}{llllll}
$\sqrt{s}$ (GeV) & $E_T^{min}$ (GeV) 
& $F^{gap}_{cem}$ (\%) 
& $F^{gap}_{exp}$ (\%) & $S_p$ (\%) 
& $F^{gap}_{cem} \times \bar{S}_p$ (\%)  
\\ \hline
1800 & 30 & 2.91 & $0.94\pm0.13$ (D\O ) & $32.3\pm4.5 $ & $1.00\pm0.10$ 
\\
1800 & 20 & 2.49 & $1.13\pm0.16$ (CDF)  & $45.4\pm6.4 $ & $0.85\pm0.08$ 
\\
1800 & 12 & 2.24 & $0.54\pm0.17$ (D\O ) & $24.1\pm7.6 $ & $0.77\pm0.07$ 
\\ \hline 
630  & 12 & 2.97 & $1.85\pm0.38$ (D\O ) & $62.3\pm12.8$ & $1.94\pm0.36$ 
\\
630  &  8 & 2.55 & $2.3 \pm1.0 $ (CDF)  & $90.2\pm39.2$ & $1.67\pm0.31$
\end{tabular}
\vskip 12pt
\caption[]
{Gap frequencies and survival probabilities. The average
survival probabilities are $\bar S_p(1800)=34.4\pm3.3$\%,
and $\bar S_p(630)=65.4\pm12.1$\%. Theoretical uncertainties 
are not included.}
\label{tab:surv}
\end{table}
\newpage


\vspace{.75in}

\vspace{.75in}

\vspace{.75in}

\vspace{.75in}

\vspace{.75in}

\end{document}